\documentclass[a4paper]{jpconf}
\usepackage{graphicx}

\usepackage{hyperref}
\input{sty/panda_commands}
\usepackage{url}
\usepackage{upgreek}

\def\Pgp{\ifmmode\mathrm{p}
         \else$\mathrm{p}$\fi}
\def\Pagp{\ifmmode\mathrm{\overline{p}}
         \else$\mathrm{\overline{p}}$\fi}
\def\Pgn{\ifmmode\mathrm{n}
         \else$\mathrm{n}$\fi}
\def\Pagn{\ifmmode\mathrm{\overline{n}}
         \else$\mathrm{\overline{n}}$\fi}
\def\Pp{\ifmmode\mathrm{p}
         \else$\mathrm{p}$\fi}
\def\Pap{\ifmmode\mathrm{\overline{p}}
         \else$\mathrm{\overline{p}}$\fi}
\def\PaB{\ifmmode\mathrm{\overline{B}}
         \else$\mathrm{\overline{B}}$\fi}
\def\PB{\ifmmode\mathrm{{B}}
         \else$\mathrm{{B}}$\fi}

\def\Pn{\ifmmode\mathrm{n}
         \else$\mathrm{n}$\fi}
\def\Pan{\ifmmode\mathrm{\overline{n}}
         \else$\mathrm{\overline{p}}$\fi}
\def\Py{\ifmmode\mathrm{Y}
         \else$\mathrm{Y}$\fi}
\def\Pay{\ifmmode\mathrm{\overline{Y}}
         \else$\mathrm{\overline{Y}}$\fi}

\def\PgL{\ifmmode\mathrm{\Lambda}
          \else$\mathrm{\Lambda}$\fi}
\def\PagL{\ifmmode\mathrm{\overline{\Lambda}}
            \else$\mathrm{\overline{\Lambda}}$\fi}
\def\PgS{\ifmmode\mathrm{\Sigma}
          \else$\mathrm{\Sigma}$\fi}
\def\PagS{\ifmmode\mathrm{\overline{\Sigma}}
            \else$\mathrm{\overline{\Sigma}}$\fi}
\def\PgSp{\ifmmode\mathrm{\Sigma^+}
          \else$\mathrm{\Sigma^+}$\fi}
\def\PagSp{\ifmmode\mathrm{\overline{\Sigma^+}}
            \else$\mathrm{\overline{\Sigma^+}}$\fi}
\def\PgSm{\ifmmode\mathrm{\Sigma^-}
          \else$\mathrm{\Sigma^-}$\fi}
\def\PagSm{\ifmmode\mathrm{\overline{\Sigma^-}}
            \else$\mathrm{\overline{\Sigma^-}}$\fi}
\def\PgSn{\ifmmode\mathrm{{\Sigma}^0}
            \else$\mathrm{{\Sigma}^0}$\fi}
\def\PagSn{\ifmmode\mathrm{\overline{\Sigma}^0}
            \else$\mathrm{\overline{\Sigma}^0}$\fi}
\def\PgX{\ifmmode\mathrm{\Xi}
          \else$\mathrm{\Xi}$\fi}
\def\PagX{\ifmmode\mathrm{\overline{\Xi}}
            \else$\mathrm{\overline{\Xi}}$\fi}
\def\PXmaXp{\ifmmode\mathrm{\Xi^-}\mathrm{\overline{\Xi}}\mathrm{^+}
          \else$\mathrm{\Xi^-}\mathrm{\overline{\Xi}}\mathrm{^+}$\fi}
\def\PXaX{\ifmmode\mathrm{\Xi}\mathrm{\overline{\Xi}}
          \else$\mathrm{\Xi}\mathrm{\overline{\Xi}}$\fi}

\def\PgXm{\ifmmode\mathrm{\Xi^-}
          \else$\mathrm{\Xi^-}$\fi}
\def\PagXm{\ifmmode\mathrm{\overline{\Xi^-}}
            \else$\mathrm{\overline{\Xi^-}}$\fi}
\def\PagXp{\ifmmode\mathrm{\overline{\Xi}}\mathrm{^+}
            \else$\mathrm{\overline{\Xi}}\mathrm{^+}$\fi}
 \def\PagXn{\ifmmode\mathrm{\overline{\Xi}}\mathrm{^0}
            \else$\mathrm{\overline{\Xi}}\mathrm{^0}$\fi}
\def\PgO{\ifmmode\mathrm{\Omega}
          \else$\mathrm{\Omega}$\fi}
\def\PagO{\ifmmode\mathrm{\overline{\Omega}}
            \else$\mathrm{\overline{\Omega}}$\fi}
\def\PgOm{\ifmmode\mathrm{\Omega^-}
          \else$\mathrm{\Omega^-}$\fi}
\def\PagOm{\ifmmode\mathrm{\overline{\Omega^-}}
            \else$\mathrm{\overline{\Omega^-}}$\fi}
\def\PgOp{\ifmmode\mathrm{\Omega^+}
          \else$\mathrm{\Omega^+}$\fi}
\def\PagOp{\ifmmode\mathrm{{\overline{\Omega}}\mathrm{^+}}
            \else$\mathrm{\overline{\Omega}}\mathrm{^+}$\fi}

\def\PgLc{\ifmmode\mathrm{\Lambda_c}
          \else$\mathrm{\Lambda_c}$\fi}
\def\PagLc{\ifmmode\mathrm{\overline{\Lambda}_c}
            \else$\mathrm{\overline{\Lambda}_c}$\fi}

\def\PgD{\ifmmode\mathrm{D}
          \else$\mathrm{D}$\fi}
\def\PagD{\ifmmode\mathrm{\overline{D}}
            \else$\mathrm{\overline{D}}$\fi}

\def\PgPi{\ifmmode\mathrm{\pi}
          \else$\mathrm{\pi}$\fi}
\def\PagPi{\ifmmode\mathrm{\overline{\pi}}
            \else$\mathrm{\overline{\pi}}$\fi}
\def\pandas{\ifmmode\mathrm{\overline{\textsc{P}}{\textsc{anda}}}
            \else$\mathrm{\overline{\textsc{P}}{\textsc{anda}}}$\fi}

\def\Kp{\ifmmode\mathrm{K^+}
          \else$\mathrm{K^+}$\fi}
\def\Km{\ifmmode\mathrm{K^-}
          \else$\mathrm{K^-}$\fi}

\usepackage[version=3]{mhchem}
\usepackage{wrapfig}

\usepackage[symbol]{footmisc}
\renewcommand{\thefootnote}{\fnsymbol{footnote}}

\begin{document}
\title{High accuracy synchrotron radiation interferometry with relativistic electrons}

\author{P~Klag$^1$, P~Achenbach$^{1,2}$, 
T~Akiyama$^3$, R~B\"ohm$^4$,
M~O~Distler$^1$, L~Doria$^1$, P~Eckert$^1$, A~Esser$^1$, J~Geratz$^1$,
T~Gogami$^5$, C~Helmel$^1$, P~Herrmann$^{1,6}$, M~Hoek$^1$,
M~Kaneta$^3$, Y~Konishi$^3$, R~Kino$^3$,
W~Lauth$^1$, H~Merkel$^1$, M~Mizuno$^3$, U~M\"uller$^1$, S~Nagao$^7$,
S~N~Nakamura$^{3,7}$, K~Okuyama$^3$, J~Pochodzalla$^{1}$, B~S~Schlimme$^1$,
C~Sfienti$^1$, T~Shao$^{1,8}$, M~Steinen$^9$, S~Stengel$^1$, M~Thiel$^1$, Y~Toyama$^{10}$}

\address{$^1$ Institute for Nuclear Physics, Johannes Gutenberg University, 55099 Mainz, Germany}
\address{$^2$ Thomas Jefferson National Accelerator Facility (JLab), Newport News, Virginia 23606, USA}
\address{$^3$ Graduate School of Science, Tohoku University, Sendai, Miyagi 980-8578, Japan}
\address{$^4$ FAIR, Planckstr. 1, 64291 Darmstadt, Germany}
\address{$^5$ Graduate School of Science, Kyoto University, Kyoto 606-8502, Japan}
\address{$^6$ GSI Helmholtzzentrum für Schwerionenforschung GmbH, 64291 Darmstadt, Germany}
\address{$^7$ Graduate School of Science, The University of Tokyo, Tokyo 113-0033, Japan}
\address{$^8$ Key Laboratory of Nuclear Physics and Ion-beam Application (MOE), Institute of Modern Physics, Fudan University, Shanghai 200433, China}
\address{$^9$ Helmholtz Institute Mainz (HIM), GSI Helmholtzzentrum für Schwerionenforschung, Johannes Gutenberg University, 55099 Mainz, Germany}
\address{$^{10}$College of Engineering, Chubu University, Kasugai, Aichi 487-8501, Japan}

\ead{pklag02@uni-mainz.de, pochodza@uni-mainz.de}

\begin{abstract}
A high-precision hypernuclear experiment has been performed at the Mainz Microtron (MAMI) to determine the hypertriton $\Lambda$ binding energy via decay-pion spectroscopy. 
A key element of this measurement is an accurate calibration of the magnetic spectrometers with the MAMI beam. For such an absolute calibration with small statistical and systematic uncertainties the undulator light interference method will be applied. In this contribution the basic principle of this method is discussed and the analysis status of the measured synchrotron radiation spectra is presented. 
\end{abstract}

\section{Motivation and background}
\begin{figure}[t]
\begin{minipage}{0.49\textwidth}
\includegraphics[width=1.0\textwidth]{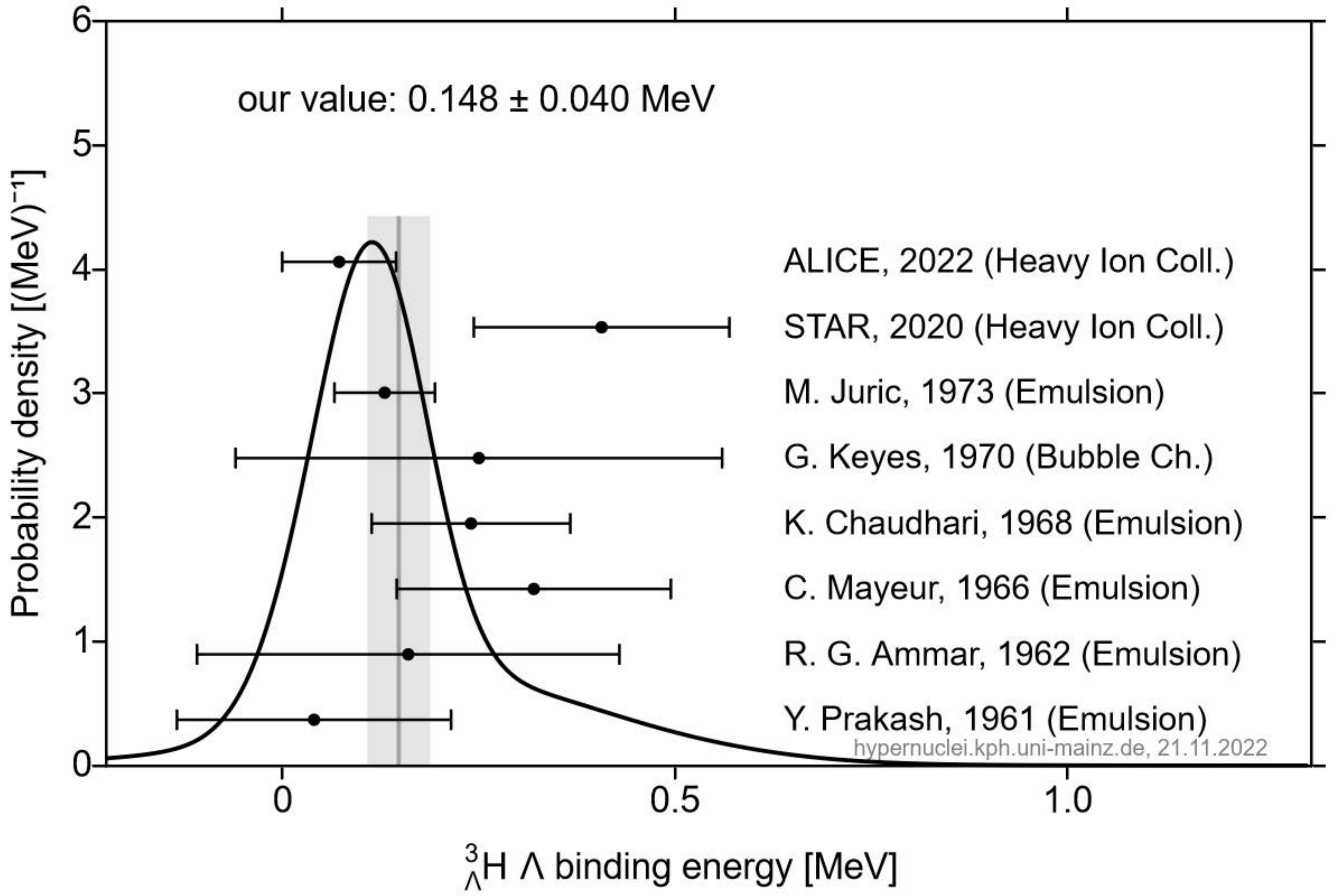}
 \caption{Ideogram of the world data set on $\Lambda$ binding energy measurements for the {\HHH}~\cite{HypernuclearDataBase}. The error bars include statistical and systematic uncertainties.
 }
 \vspace{1.0cm}
 \caption{Setup for the hypernuclear fragmentation experiment at MAMI. The electron beam (coming from the back) was steered through a beam transport-line magnetic chicane and was incident on the $^9$Be target. The \kaos\ spectrometer (purple) was installed at 0$^\circ$ forward angle as a kaon tagger. The
spectrometers SpekA and SpekC were used as decay-pion spectrometers. Fig. from Ref.~\cite{Schulz2015PhDthesis}.}
 \label{fig:HL_binding}
 \end{minipage}
 \hfill
 \begin{minipage}{0.47\textwidth}
 \includegraphics[width=1.0\textwidth]{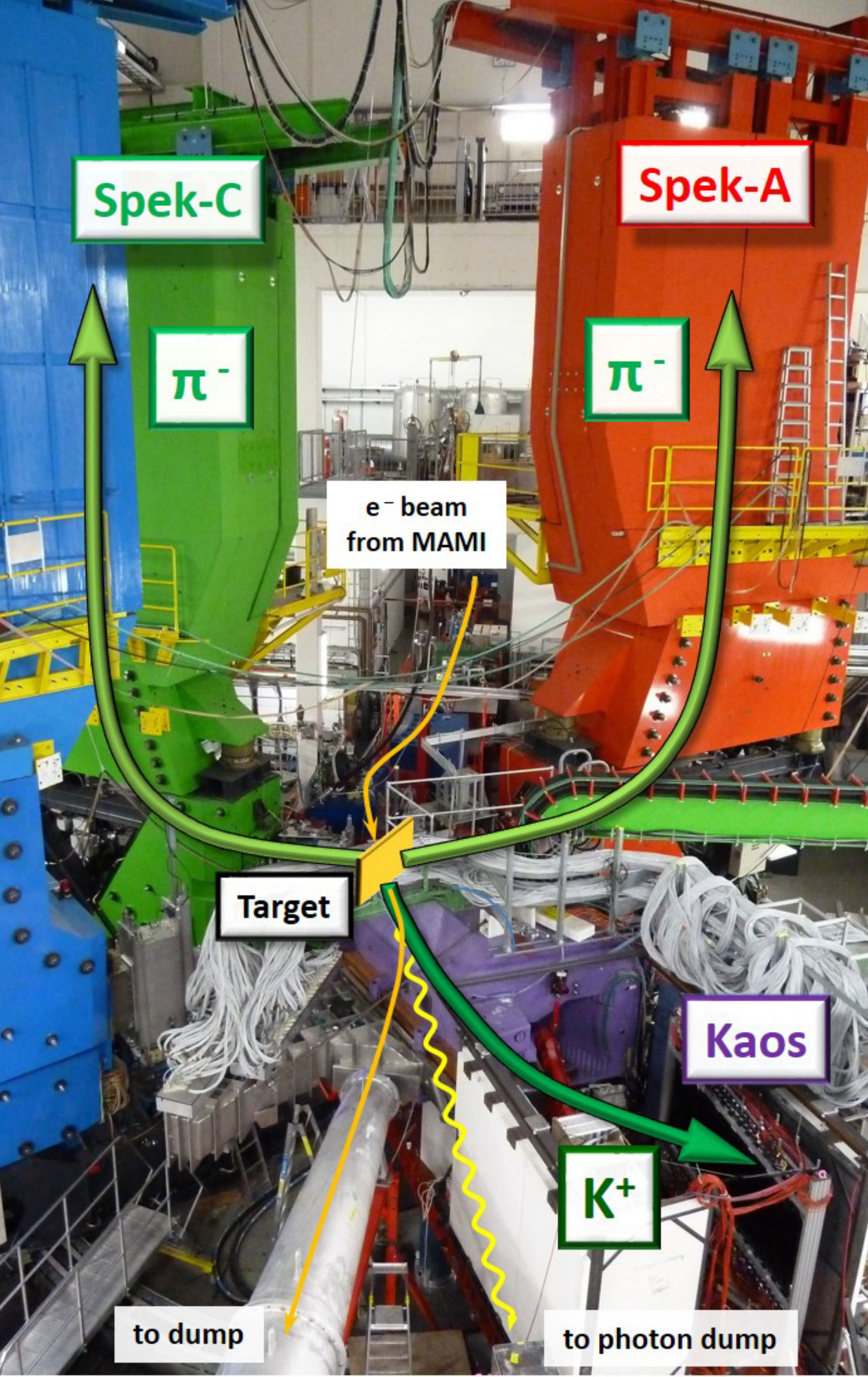}
\end{minipage}
\label{fig:Setup2012}
\end{figure}

One of the most intriguing questions in physics is: what do we find in the inner core of dense stellar objects such as neutron stars? There has been a wide consensus in nearly all theoretical approaches for neutron star matter that hyperons may appear in these cores at densities of about twice the nuclear saturation density.
However, the determination of the nuclear equation of state at such densities remains one of the biggest challenges in nuclear physics even after many decades of research. Hadrons with strangeness embedded in the nuclear environment, hypernuclei or strange atoms, are the only available tool to approach the many-body aspect of the three-flavor strong interaction.

Light hypernuclei are particularly interesting since not only phenomenological models but also ab initio studies based on chiral effective field theory including three-baryon forces can be performed. Furthermore, lattice quantum chromodynamics calculations are within reach for such nuclei. A profound understanding of the lightest hypernuclei, the hypertriton {\HHH}, is therefore a cornerstone for any strong interaction theory dealing with strange baryons.

For almost 50 years, its most precise binding energy value is given by $B_{\Lambda} = 130 \pm 50$\,keV, averaged and compiled by Juri\v{c} et al.\ from emulsion experiments~\cite{JURIC19731, BOHM1968511}. Just recently, two new values became available, one by the STAR Collaboration~\cite{STARmass2020},
$B_{\Lambda} = 406 \pm 120\, \mathrm{(stat.)} \pm 110\, \mathrm{(syst.)}$\,keV,
and one by the ALICE Collaboration~\cite{ALICE:2022rib},
$B_{\Lambda} = 72 \pm 63\, \mathrm{(stat.)} \pm 36\, \mathrm{(syst.)}$\,keV.
Remarkably, the STAR value is about 5 times larger than the one from ALICE and they differ by about two standard deviations. The STAR value also seems to be in tension with the emulsion value.
This data situation along with other earlier measurements is visualized in an ideogram in Fig.~\ref{fig:HL_binding}, obtained from the Chart of Hypernuclides~\cite{HypernuclearDataBase}. An average value of
$\overline{B_{\Lambda}} = 148 \pm 40$\,keV is computed with a relative error of still more than 25\,\%.
In the coming years, several experiments will produce more precise and accurate data for the hypertriton. The WASA-FRS Collaboration at GSI/FAIR, the E73/77 collaboration at J-PARC \cite{P73-JPARCProposal,E77-proposal},
ALICE during LHC Run3 and Run4, and an experiment at ELPH at Tohoku University will improve our knowledge on the lifetime of {\HHH}. New measurements of the {\HHH} binding energy are expected from experiments at the Mainz Microtron (MAMI),
JLab \cite{JLAB-C12-19-002-proposal}, from ALICE, and from the J-PARC E07 emulsion experiment. These many attempts attest the importance of the hypertriton as a benchmark for hypernuclear structure calculations.

In 2012, the first high-resolution spectroscopy of pions from decays
of stopped $_{\Lambda}^{4}$H hypernuclei was performed by the A1
Collaboration at MAMI~\cite{PhysRevLett.114.232501}. $_{\Lambda}^{4}$H hypernuclei were produced in a multi-step strangeness production, nuclear fragmentation, and pionic weak decay
reaction:
\begin{equation}
  \rm{^{9}Be}(e,e'K^+)\rm{{}^9_\Lambda Li}^* \quad \rightarrow \quad
  X + \rm{{}^4_\Lambda H}; \hspace{1cm}
  \rm{{}^4_\Lambda H}  \rightarrow  \rm{{}^4He} + \pi^-
\end{equation}
Pions were detected alternatively in one of
the two spectrometers SpekA or SpekC in coincidence with positive
kaons tagged in the \kaos\ spectrometer.

\begin{figure}
    \begin{minipage}{0.48\textwidth}
    \includegraphics[width=1.0\textwidth]{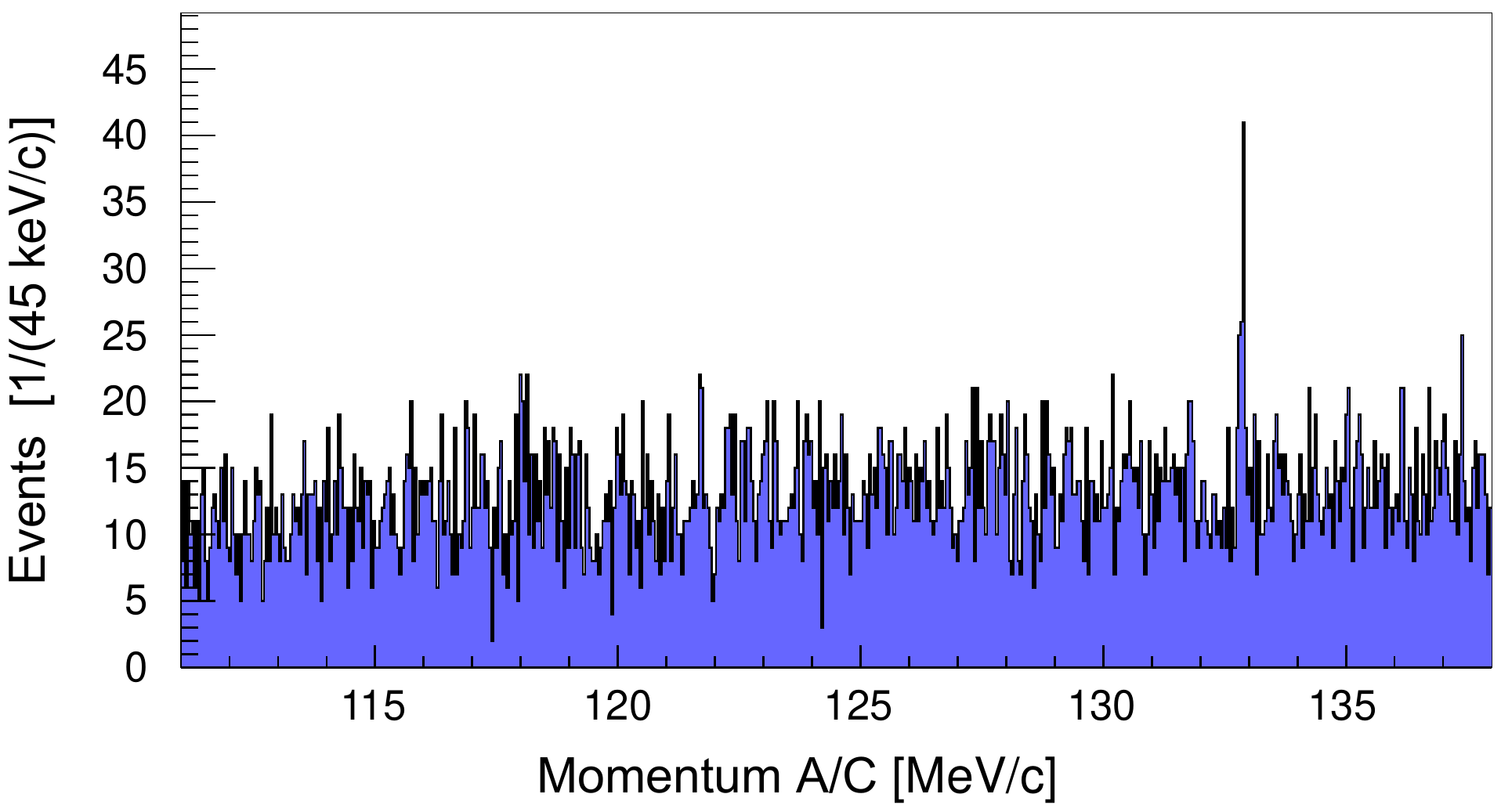}
    \caption{Momentum spectrum for strangeness tagged pions from the 2014 measurement at MAMI~\cite{Schulz2016149}. Mono-energetic decay-pions of {\HH} were observed at $\sim$133\,MeV$\!/c$. A  signal from two-body decays of stopped {\HHH} was not found at the expected momentum of $\sim$114\,MeV$\!/c$.}
    \label{fig:result2016}
\end{minipage}
\hfill
\begin{minipage}{0.5\textwidth}
\centering
    \includegraphics[width=0.98\textwidth]{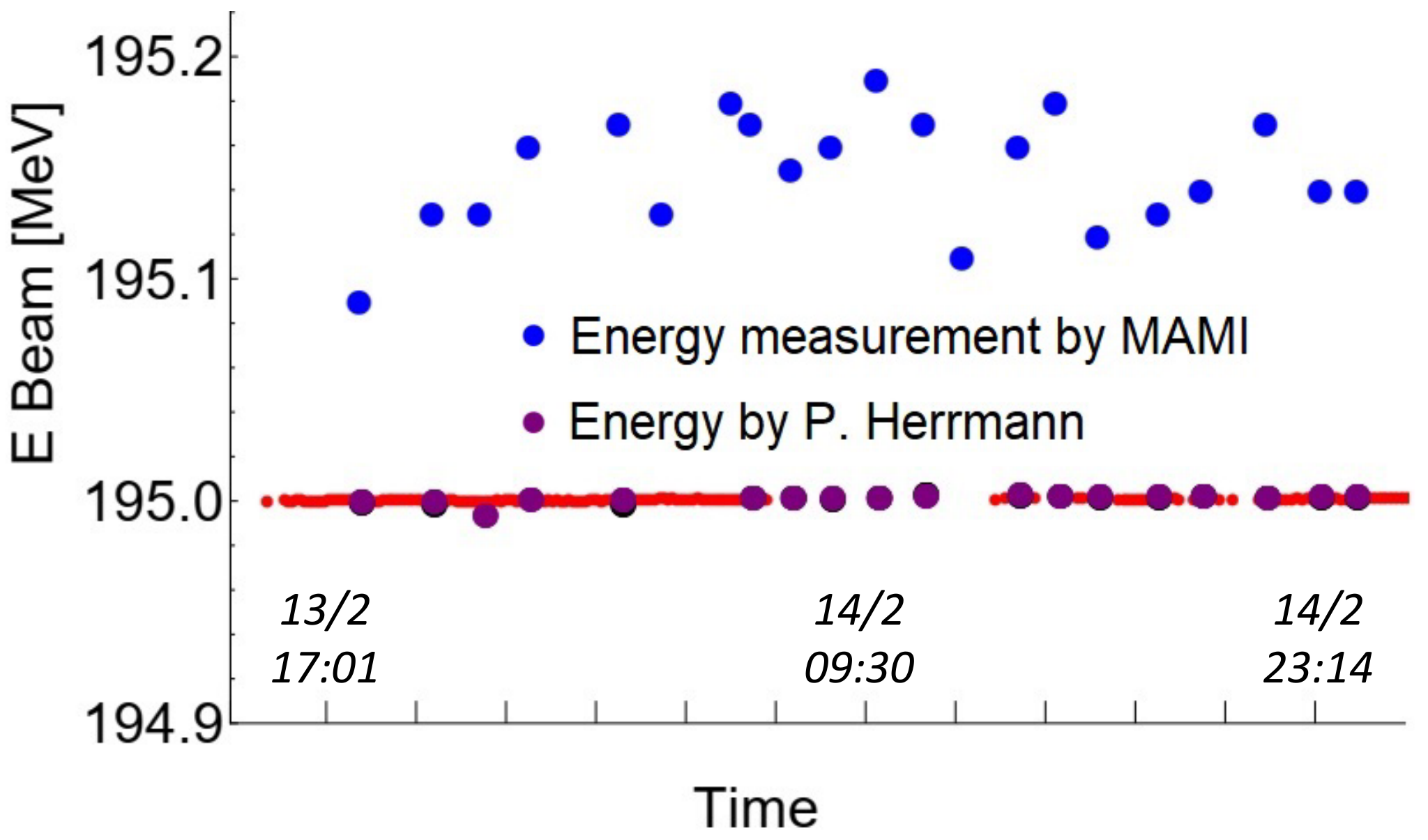}
    \caption{MAMI beam energy measured over a period of about 30 hours \cite{Herrmann2021PhDthesis}. The blue points show MAMI measurements from the position of the beam within the Race Track Microton RTM3~\cite{KAISER2008159}.
    The purple points are derived from a precise measurement of the deflection in a dipole magnet of the A1 beamline.}
    \label{fig:energy2016}
\end{minipage}
\end{figure}
In the year 2014 the experiment at MAMI was continued with improved control of systematic uncertainties, additional background suppression, and higher luminosity. A detailed description of the
2014 experiment can be found in Refs.~\cite{Schulz2015PhDthesis,Schulz2016149}.
Fig.~\ref{fig:result2016} shows a representative momentum spectrum, where decay pions of {\HH} are clearly visible at around 133\,MeV$\!/c$, resulting in a binding energy of
$B_{\Lambda} = 2.157 \pm 0.005\, \mathrm{(stat.)} \pm 0.077\, \mathrm{(syst.)}$\,MeV.
The systematic error was strongly dominating because the spectrometer calibration was limited by the available MAMI beam energy measurement with an accuracy of 160\,keV, see blue points in Fig.~\ref{fig:energy2016}. However, the beam spread and instability are known to be much smaller. This could be verified by a precise measurement of the deflection in a dipole magnet of the A1 beamline as shown by the purple points in Fig.~\ref{fig:energy2016} \cite{Herrmann2021PhDthesis}. In this study, the stability of the dipole magnet was controlled by an NMR probe whose readout (on arbitrary scale) is shown by the small red points.

As the data situation for {\HHH}  is clearly not satisfactory, a new high-precision experiment via decay-pion spectroscopy has been commissioned at MAMI with the goal to reach a 20\,keV total error in the binding energy~\cite{Achenbach:2018WpProc, Eckert:2022G6}. This will be made possible by measuring the absolute value of the beam energy with a novel undulator light interference method \cite{KLAG2018147}.

\section{Basic concept of the measurement}

\begin{figure}[tb]
    \includegraphics[width=1.0\textwidth]{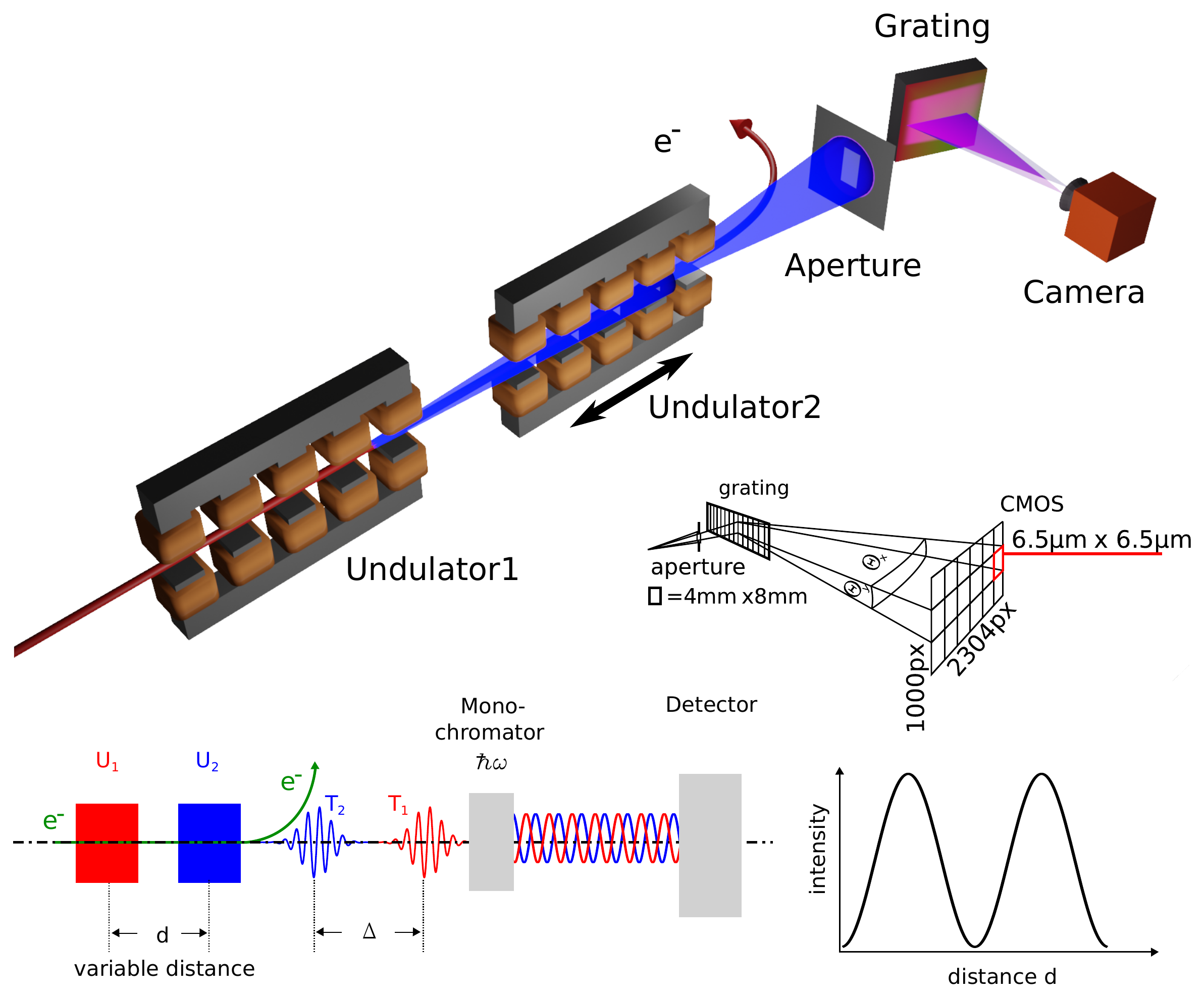}
    \caption{Schematic layout (not to scale) of the experimental setup. Relativistic electrons pass through two undulators separated by a variable distance, thus generating coherent pairs of electromagnetic wave packets as illustrated in the lower part. The insert illustrates the coordinate system of the optical system.
    The x-axis describes the dispersive plane and the y-axis denotes the vertical off-axis direction.
    }
    \label{fig:SetupUnd}
\end{figure}

The method is based on the analysis of the intensity oscillation length in the synchrotron spectrum from two collinear sources, which are realized by two undulators and the electron beam of MAMI \cite{PhysRevLett.80.5473}.
The schematic layout of the experimental setup is displayed in Fig.~\ref{fig:SetupUnd}. Relativistic electrons traverse two identical undulators thus generating coherent pairs of electromagnetic wave packets as illustrated in the lower part. The time delay between the two wave packets can be controlled by changing the distance $d$ between the two undulators.

After passing through an aperture with an opening of
$a_\mathrm{y} \times a_\mathrm{x} = \pm 4\,\mathrm{mm} \times \pm 2$\,mm, the radiation is dispersed according to its wavelength by a reflective grating spectrometer \cite{JY1993} and detected by a CMOS camera \cite{HAMAMATSU:2019}. The CMOS chip has a pixel size of 6.5\,$\mu$m $\times$ 6.5\,$\mu$m and a resolution of 2304 $\times$ 2304 pixels. In this setup one pixel covers a wavelength interval of $\Delta\lambda_\mathrm{rad}$ = 0.0048\,nm.

The intensity for a selected wavelength $\lambda_\mathrm{rad}$ varies periodically with the distance ${\Delta}d$ between the two undulators as illustrated in the lower part in Fig.~\ref{fig:SetupUnd}. Here, ${\Delta}d$ is measured with respect to the closest position of the two undulators. The oscillation length $\lambda_\mathrm{osc}$ is directly related to the Lorentz factor of the electron beam and the wavelength of the radiation
\begin{equation}
  \gamma^2 =\frac{1}{2}\frac{\lambda_\mathrm{osc}}{\lambda_\mathrm{rad}}.
\label{eq:gam}
\end{equation}
Thus the energy determination is reduced to a relative distance measurement in the decimeter range and the spectroscopy in a narrow optical wavelength band.

Eq.~\ref{eq:gam} only holds in the limit of a precise alignment between the electron beam and the optical axis \cite{PhysRevLett.80.5473}. For off-axis observation angles $\theta \neq 0$, Eq.~\ref{eq:gam} is modified according to
\begin{equation}
  \gamma^2 =\frac{\lambda_\mathrm{osc}}{\lambda_{\mathrm{osc}}\cdot\theta^2-2\lambda_\mathrm{rad}}.
\end{equation}
This causes a bending of the interference spectrum in the non-dispersive direction.

\section{Measurements and analysis of synchrotron radiation spectra}

\begin{figure}[bt]
\begin{minipage}[c]{0.49\textwidth}
 \includegraphics[width=0.95\textwidth]{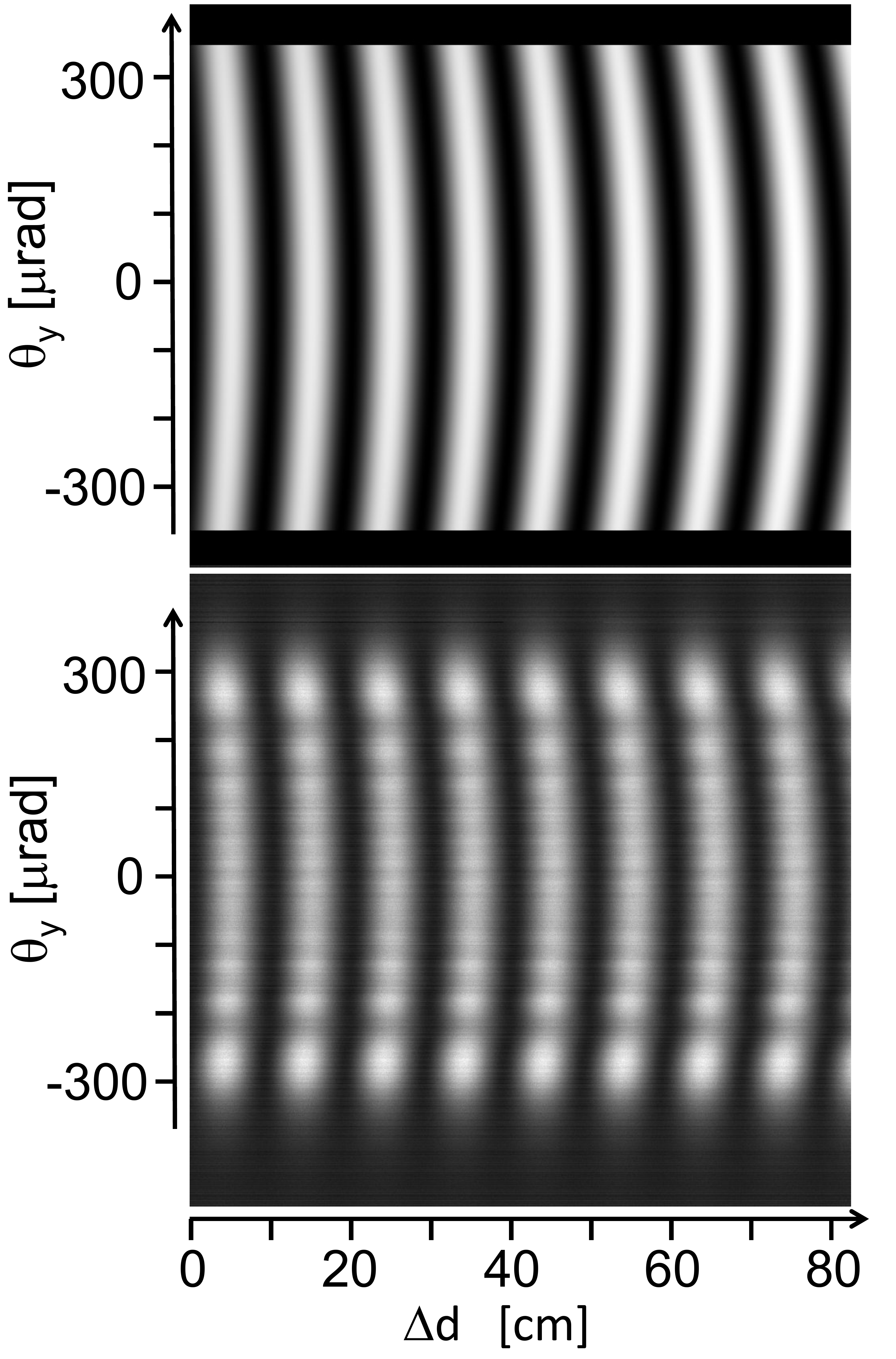}
    \caption{Top: Theoretical interference signal for a wavelength of 404.58\,nm with varying distance ${\Delta}d$ between the two undulators and its dependence on the vertical off-axis angle $\theta_\mathrm{y}$. The calculations do not include diffraction. Bottom: Measured interference signal for the same wavelength. Because of the finite size of the aperture $a_\mathrm{y}\,=\,\pm$4\,mm a characteristic diffraction pattern is present.
    }
    \label{fig:TheoExp}
\end{minipage}
\hfill
\begin{minipage}[c]{0.49\textwidth}
 \includegraphics[width=1.0\textwidth]{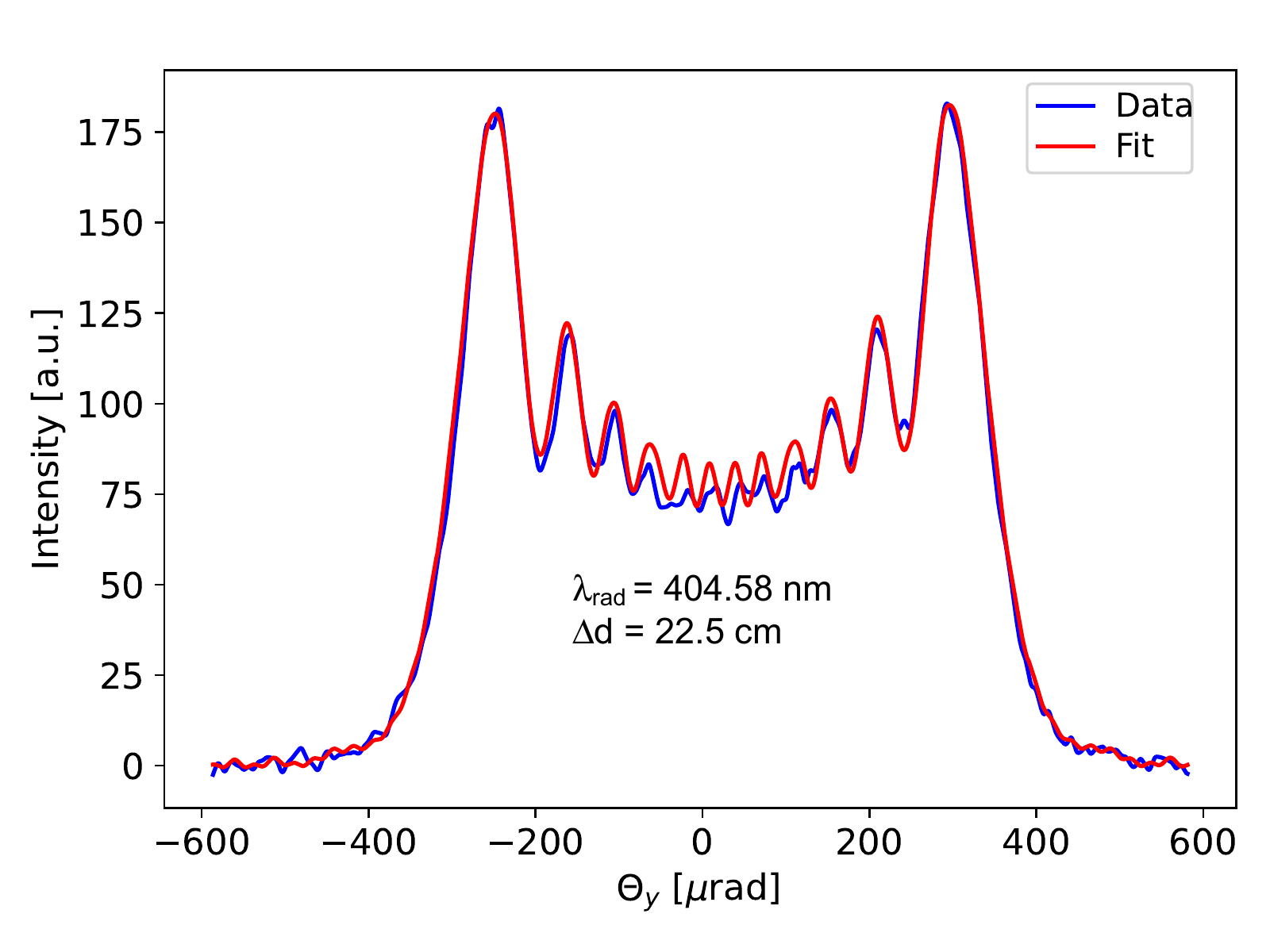}
 \includegraphics[width=1.0\textwidth]{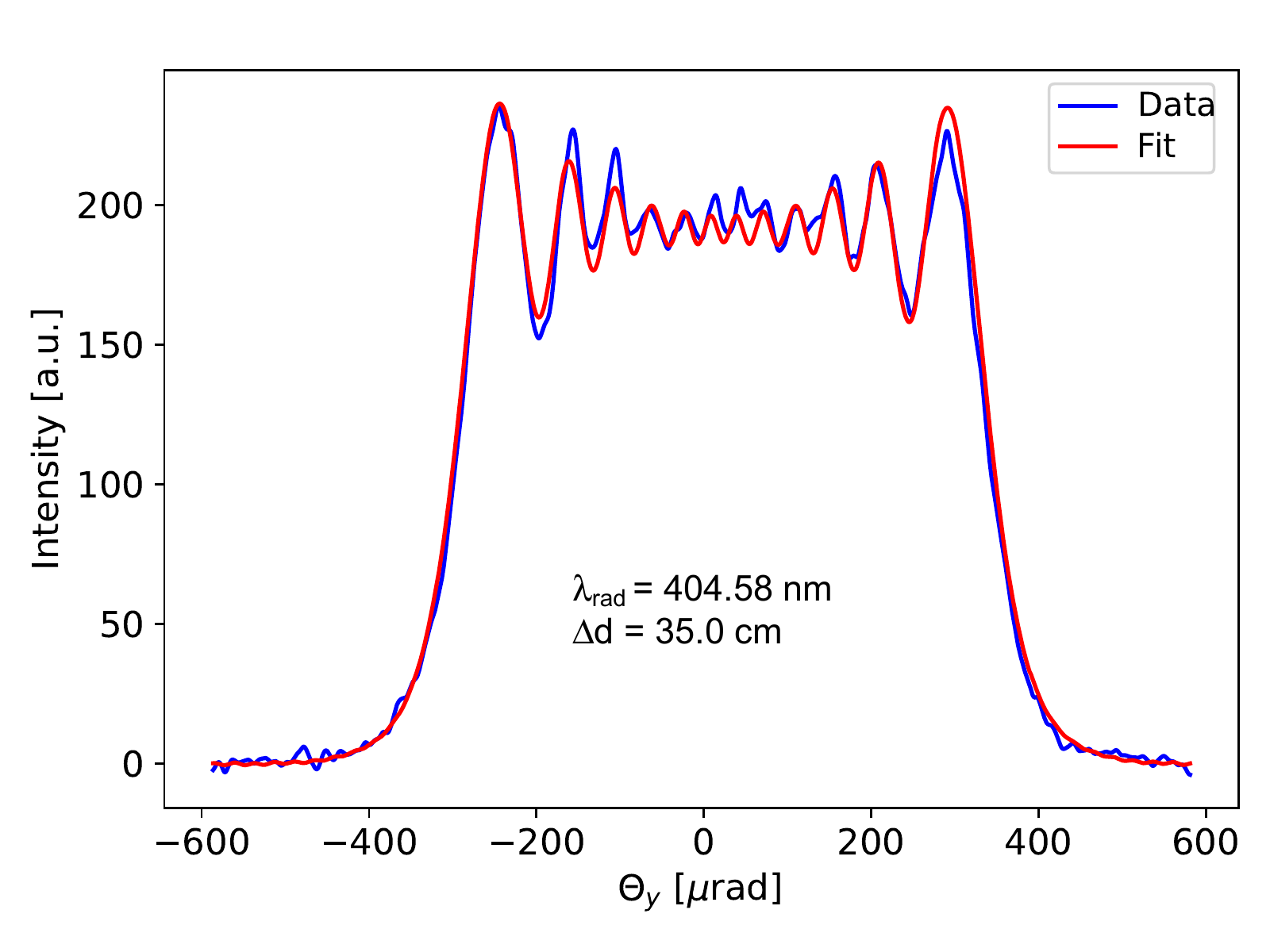}
    \caption{Two representative examples for observed radiation intensities as a function of the vertical off-axis angle $\theta_\mathrm{y}$ for undulator distances of ${\Delta}d$ = 22.5\,cm (top) and 35.0\,cm (bottom).
    A wavelength bin at $\lambda_\mathrm{rad}$ = 404.58\,nm was selected. For each wavelength bin 825 spectra were recorded and analyzed.
    }
    \label{fig:fit01}
\end{minipage}
\end{figure}

The upper part of Fig.~\ref{fig:TheoExp} shows the theoretical variation of the interference signal with varying distance ${\Delta}d$ between the two undulators and vertical off-axis angle $\theta_\mathrm{y}$, if no other limitations would apply. However, the synchrotron radiation is collimated by the finite size of the aperture with $a_y=\pm$4\,mm. As a consequence, a characteristic diffraction pattern is present in the experimentally measured spectrum as shown in the lower part of Fig.~\ref{fig:TheoExp}.
This structure proves to be very useful for the determination the true $\theta_\mathrm{y}$ = 0° direction during the fitting procedure.
In the dispersive direction $\theta_\mathrm{x}$, the aperture is not relevant, since the grating separates wavelengths independent of the position of the incident radiation.

\begin{figure}[tb]
\begin{center}
  \includegraphics[width=\textwidth]{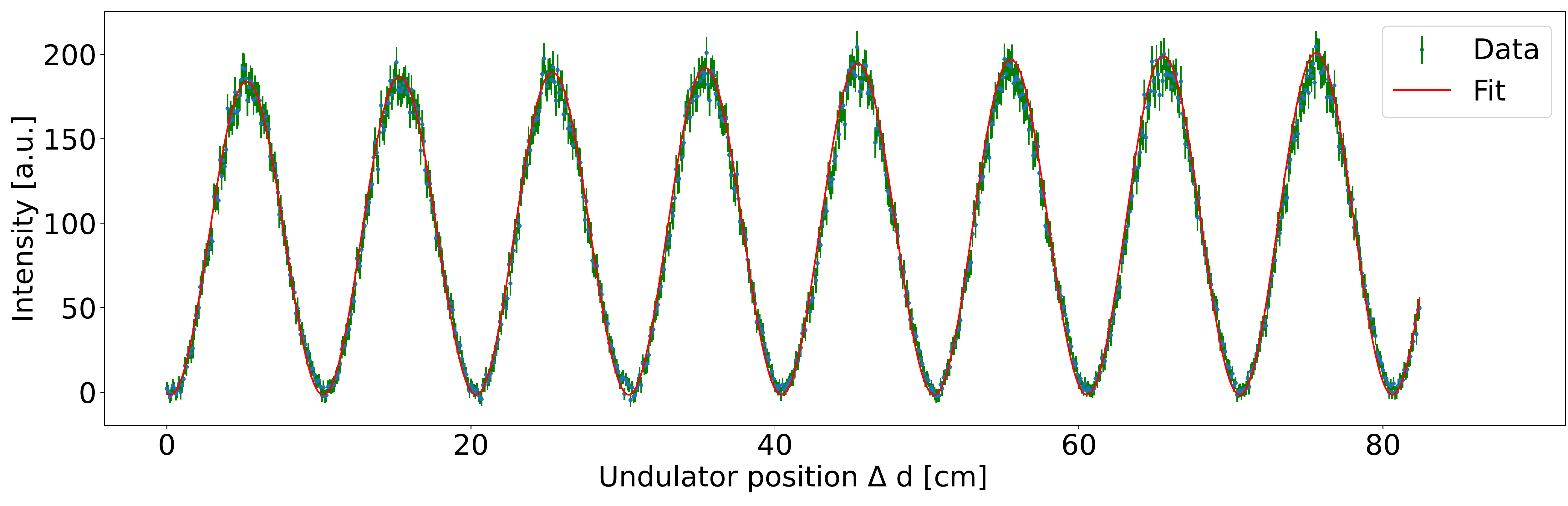}
\end{center}
\caption{Intensity for $\theta_\mathrm{y}$ = 0 and $\lambda_\mathrm{rad}$ = 404.58\,nm as a function of the  undulator distance ${\Delta}d$, c.f.\ lower part of Fig.~\ref{fig:SetupUnd}. The blue points show the measured intensity. Statistical errors are indicated in green. The red curve shows the fit discussed in the text.}
\label{fig:oscillation}

\end{figure}

Prior to the measurement, the beamline was aligned optically with a theodolite and a diode laser system at a wavelength of $\lambda_\mathrm{las}$ = 405\,nm. For the energy calibration of the combination of the grating spectrometer and the CMOS camera, a mercury lamp was inserted upstream of the stationary first undulator. The Hg lamp has two dominant, well known lines within the acceptable wavelength band at 404.6565\,nm and 407.7837\,nm~\cite{NIST_ASD}.
In practice, the radiation from the Hg lamp, the undulators, and the laser have differing divergences: the Hg lamp emits approximately isotropically, whereas the undulator radiation is forward peaked with a typical angle of $1/\gamma \sim$\, $0.15\textrm{°}$. However, both radiation fields are collimated by the rectangular aperture to about $\theta_x <$\,
$0.01\textrm{°}$ leading to a similar wavelength resolution. On the contrary, the divergence of the laser beam is in the range of $0.02\textrm{°} - 0.03\textrm{°}$ and thus slightly more focused. The measured width of the 407.78\,nm Hg line was 4.2 pixels (FWHM), whereas the laser line was only 3.0 pixels (FWHM) wide.

In the analysis of the experimental data, 5 pixel columns of the camera were combined to a wavelength bin in order to improve the signal-to-noise ratio. Such a bin corresponds to a relative wavelength window of
\begin{equation}
  \frac{\Delta\lambda_\mathrm{rad}}{\lambda_\mathrm{rad}} = \frac{0.024\,\mathrm{nm}}{405\,\mathrm{nm}} = 6\cdot10^{-5}.
\end{equation}
This bin is sufficiently small to have no sizable effect on the determination of the 
MAMI beam energy after a proper calibration of the $\lambda_\mathrm{rad}$ scale. 

Fig.~\ref{fig:fit01} shows two examples of the measured interference spectra for a wavelength of $\lambda_\mathrm{rad}$ = 404.58\,nm at ${\Delta}d$ = 22.5\,cm and 35.0\,cm, respectively.  For each selected $\lambda_\mathrm{rad}$ bin, 825 positions ${\Delta}d$ of the movable undulator were analyzed.  
The blue data points in Fig.~\ref{fig:oscillation} show the observed oscillation of the intensity for $\theta_\mathrm{y}$ = 0 and $\lambda_\mathrm{rad}$ = 404.58\,nm as a function of the  undulator distance ${\Delta}d$, c.f.\ lower part of Fig.~\ref{fig:SetupUnd}. In total, 460 different wavelength bins were
measured. For every set of data, a background image has been taken which is subtracted. Even under optimal conditions offsets remain. 

For each given wavelength $\lambda_\mathrm{rad}$, the interference spectra for 825 measured undulator positions ${\Delta}d$ were fitted simultaneously with only 7 parameters. The first five parameters are directly related to the experimental setup:
\begin{enumerate}
\item
The absolute intensity, which is, however, not relevant for the MAMI energy determination (c.f.\ Eq.~\ref{eq:gam}).
\item
The phase factor of the oscillation, which reflects the effective offset of ${\Delta}d$.
\item
The effective distance between the aperture and the camera of approximately 5.45\,m.
\item
A displacement of the source with respect to the center of the aperture.
\item
An elevation parameter, which describes the increase of the intensity when moving the second undulator closer towards the aperture.
\item
A parameter, which describes the residual intensity on the camera independent of the undulator radiation.
\item
The relativistic $\gamma$ factor, which determines the MAMI beam energy.
\end{enumerate}

Figs.~\ref{fig:fit01} and \ref{fig:oscillation} include  
the results of the fitting procedure for $\lambda_\mathrm{rad}$ = 404.58\,nm.
The Fresnel pattern in the two representative $\theta_\mathrm{y}$ spectra
in Fig.~\ref{fig:fit01} is well described. Simultaneously, the oscillating interference signal for  $\theta_\mathrm{y}$ = $0\textrm{°}$ is correctly reproduced.
Presently, the analysis of all 460 different wavelength bins is being finalized. This will allow to determine the systematic uncertainty of the beam energy measurement with the projected high accuracy.

\section*{Acknowledgements}
We would like to thank the MAMI operators, technical staff, and the accelerator group for their excellent support of the experiments. This project has received funding from the European Union’s Horizon 2020 research and innovation programme under grant agreement No 824093. We acknowledge the support by Deutsche Forschungsgemeinschaft (DFG), Germany through Research Grants PO 256/7-1 and PO 256/8-1, JSPS KAKENHI grants no.18H05459 and the Graduate Program on Physics for the Universe, Tohoku University (GP-PU).

The presented data were in part collected within the framework of the PhD thesis of Pascal Klag at the Johannes Gutenberg University Mainz.

\section*{References}
\bibliography{IARD22_Klag
.bib}

\providecommand{\newblock}{}
\begin{thebibliography}{10}
\expandafter\ifx\csname url\endcsname\relax
  \def\url#1{{\tt #1}}\fi
\expandafter\ifx\csname urlprefix\endcsname\relax\def\urlprefix{URL }\fi
\providecommand{\eprint}[2][]{\url{#2}}

\bibitem{HypernuclearDataBase}
{Institut f{\"u}r Kernphysik Mainz} 2022 {Chart of Hypernuclides - Hypernuclear
  Structure and Decay Data} visited on 2022-11-21
  \urlprefix\url{https://hypernuclei.kph.uni-mainz.de}

\bibitem{Schulz2015PhDthesis}
Schulz F 2015 {\em {Pr{\"a}zisionsmessung der $\Lambda$-Separationsenergie von
  $^4_{\Lambda}$H am Mainzer Mikrotron}\/} Ph.D. thesis {Johannes
  Gutenberg-Universit{\"a}t Mainz}

\bibitem{JURIC19731}
Juric M, Bohm G, Klabuhn J, Krecker U, Wysotzki F, Coremans-Bertrand G, Sacton
  J, Wilquet G, Cantwell T, Esmael F, Montwill A, Davis D, Kielczewska D,
  Pniewski T, Tymieniecka T and Zakrzewski J 1973 {\em Nuclear Physics B\/}
  {\bf 52} 1 -- 30 ISSN 0550-3213
  \urlprefix\url{http://www.sciencedirect.com/science/article/pii/0550321373900849}

\bibitem{BOHM1968511}
Bohm G, Klabuhn J, Krecker U, Wysotski F, Coremans G, Gajewski W, Mayeur C,
  Sacton J, Vilain P, Wilquet G, O'Sullivan D, Stanley D, Davis D, Fletcher E,
  Lovell S, Roy N, Wickens J, Filipkowski A, Garbowska-Pniewska K, Pniewski T,
  Skrzypczak E, Sobczak T, Allen J, Bull V, Conway A, Fishwick A and March P
  1968 {\em Nuclear Physics B\/} {\bf 4} 511 -- 526 ISSN 0550-3213
  \urlprefix\url{http://www.sciencedirect.com/science/article/pii/0550321368901090}

\bibitem{STARmass2020}
Adam J, Adamczyk L, Adams J~R, Adkins J~K, Agakishiev G, Aggarwal M~M, Ahammed
  Z, Ajitanand N~N, Alekseev I, Alford J, Anderson D~M, Aoyama R, Aparin A,
  Arkhipkin D, Aschenauer E~C, Ashraf M~U, Attri A, Averichev G~S, Bai X,
  Bairathi V, Barish K, Behera A, Bellwied R, Bhasin A, Bhati A~K, Bhattarai P,
  Bielcik J, Bielcikova J, Bland L~C, Bordyuzhin I~G, Bouchet J, Brandenburg
  J~D, Brandin A~V, Brown D, Bryslawskyj J, Bunzarov I, Butterworth J, Caines
  H, Calder\'on de~la Barca~S\'anchez M, Campbell J~M, Cebra D, Chakaberia I,
  Chaloupka P, Chang Z, Chankova-Bunzarova N, Chatterjee A, Chattopadhyay S,
  Chen X, Chen X, Chen J~H, Cheng J, Cherney M, Christie W, Contin G, Crawford
  H~J, Das S, Dedovich T~G, Deng J, Deppner I~M, Derevschikov A~A, Didenko L,
  Dilks C, Dong X, Drachenberg J~L, Draper J~E, Dunlop J~C, Efimov L~G, Elsey
  N, Engelage J, Eppley G, Esha R, Esumi S, Evdokimov O, Ewigleben J, Eyser O,
  Fatemi R, Fazio S, Federic P, Federicova P, Fedorisin J, Feng Z, Filip P,
  Finch E, Fisyak Y, Flores C~E, Fujita J, Fulek L, Gagliardi C~A, Geurts F,
  Gibson A, Girard M, Grosnick D, Gunarathne D~S, Guo Y, Gupta A, Guryn W,
  Hamad A~I, Hamed A, Harlenderova A, Harris J~W, He L, Heppelmann S,
  Heppelmann S, Herrmann N, Hirsch A, Horvat S, Huang B, Huang T, Huang X,
  Huang H~Z, Humanic T~J, Huo P, Igo G, Jacobs W~W, Jentsch A, Jia J, Jiang K,
  Jowzaee S, Judd E~G, Kabana S, Kalinkin D, Kang K, Kapukchyan D, Kauder K, Ke
  H~W, Keane D, Kechechyan A, Khan Z, Kiko\l{}a D~P, Kim C, Kisel I, Kisiel A,
  Kochenda L, Kocmanek M, Kollegger T, Kosarzewski L~K, Kraishan A~F, Krauth L,
  Kravtsov P, Krueger K, Kulathunga N, Kumar L, Kvapil J, Kwasizur J~H, Lacey
  R, Landgraf J~M, Landry K~D, Lauret J, Lebedev A, Lednicky R, Lee J~H, Li X,
  Li W, Li Y, Li C, Lidrych J, Lin T, Lisa M~A, Liu F, Liu P, Liu Y, Liu H,
  Ljubicic T, Llope W~J, Lomnitz M, Longacre R~S, Luo X, Luo S, Ma G~L, Ma L,
  Ma R, Ma Y~G, Magdy N, Majka R, Mallick D, Margetis S, Markert C, Matis H~S,
  Mayes D, Meehan K, Mei J~C, Miller Z~W, Minaev N~G, Mioduszewski S, Mishra D,
  Mizuno S, Mohanty B, Mondal M~M, Morozov D~A, Mustafa M~K, Nasim M, Nayak
  T~K, Nelson J~M, Nemes D~B, Nie M, Nigmatkulov G, Niida T, Nogach L~V, Nonaka
  T, Nurushev S~B, Odyniec G, Ogawa A, Oh K, Okorokov V~A, Olvitt D, Page B~S,
  Pak R, Pandit Y, Panebratsev Y, Pawlik B, Pei H, Perkins C, Pluta J,
  Poniatowska K, Porter J, Posik M, Pruthi N~K, Przybycien M, Putschke J,
  Quintero A, Ramachandran S, Ray R~L, Reed R, Rehbein M~J, Ritter H~G, Roberts
  J~B, Rogachevskiy O~V, Romero J~L, Roth J~D, Ruan L, Rusnak J, Rusnakova O,
  Sahoo N~R, Sahu P~K, Salur S, Sandweiss J, Saur M, Schambach J, Schmah A~M,
  Schmidke W~B, Schmitz N, Schweid B~R, Seger J, Sergeeva M, Seto R, Seyboth P,
  Shah N, Shahaliev E, Shanmuganathan P~V, Shao M, Shen W~Q, Shi S~S, Shi Z,
  Shou Q~Y, Sichtermann E~P, Sikora R, Simko M, Singha S, Skoby M~J, Smirnov N,
  Smirnov D, Solyst W, Sorensen P, Spinka H~M, Srivastava B, Stanislaus T~D~S,
  Stewart D~J, Strikhanov M, Stringfellow B, Suaide A~A~P, Sugiura T, Sumbera
  M, Summa B, Sun Y, Sun X, Sun X~M, Surrow B, Svirida D~N, Tang A~H, Tang Z,
  Taranenko A, Tarnowsky T, Tawfik A, Th{\"a}der J, Thomas J~H, Timmins A~R,
  Tlusty D, Todoroki T, Tokarev M, Trentalange S, Tribble R~E, Tribedy P,
  Tripathy S~K, Trzeciak B~A, Tsai O~D, Ullrich T, Underwood D~G, Upsal I,
  Van~Buren G, van Nieuwenhuizen G, Vasiliev A~N, Videb\ae{}k F, Vokal S,
  Voloshin S~A, Vossen A, Wang G, Wang Y, Wang F, Wang Y, Webb G, Webb J~C, Wen
  L, Westfall G~D, Wieman H, Wissink S~W, Witt R, Wu Y, Xiao Z~G, Xie G, Xie W,
  Xu Y~F, Xu J, Xu Q~H, Xu N, Xu Z, Yang S, Yang Y, Yang C, Yang Q, Ye Z, Ye Z,
  Yi L, Yip K, Yoo I~K, Yu N, Zbroszczyk H, Zha W, Zhang Z, Zhang J, Zhang S,
  Zhang S, Zhang J, Zhang Y, Zhang X~P, Zhang J~B, Zhao J, Zhong C, Zhou L,
  Zhou C, Zhu X, Zhu Z, Zurek M and Zyzak M (STAR) 2020 {\em Nature Physics\/}
  {\bf 16} 409--412 ISSN 1745-2481
  \urlprefix\url{http://dx.doi.org/10.1038/s41567-020-0799-7}

\bibitem{ALICE:2022rib}
ALICE-Collaboration (ALICE) 2022 {Measurement of the lifetime and $\Lambda$
  separation energy of $^{3}_{\Lambda}\mathrm H$} (\textit{Preprint}
  \eprint{2209.07360})

\bibitem{P73-JPARCProposal}
Akaishi T, Asano H, X C, A C, Curceanu C, Del~Grande R, Guaraldo C, Han C,
  Hashimoto T, Iliescu M, Inoue K, Ishimoto S, Itahashi K, Iwasaki M, Ma Y,
  Miliucci M, Noumi H, Ohnishi H, Okada S, Outa H, Piscicchia K, Sakuma F, Sato
  M, Scordo A, Sirghi D, Sirghi F~Shirotori K, Suzuki S, Tanida K, Yamaga T,
  Yuan X, Zhang P, Zhang Y and Zhang H 2019 {Feasibility study for {\HHH}
  mesonic weak decay lifetime measurement with $^4$He(K$^-$, $\pi^0$){\HH}
  reaction}
  \urlprefix\url{http://j-parc.jp/researcher/Hadron/en/pac_1901/pdf/P73_2019-06.pdf}

\bibitem{E77-proposal}
Akaishi T, Asano H, X C, A C, Curceanu C, Del~Grande R, Guaraldo C, Han C,
  Hashimoto T, Iliescu M, Inoue K, Ishimoto S, Itahashi K, Iwasaki M, Ma Y,
  Miliucci M, Noumi H, Ohnishi H, Okada S, Outa H, Piscicchia K, Sakuma F, Sato
  M, Scordo A, Sirghi D, Sirghi F~Shirotori K, Suzuki S, Tanida K, Yamaga T,
  Yuan X, Zhang P, Zhang Y and Zhang H {J-PARC E77 experiment: Feasibility
  study for {\HHH} mesonic weak decay lifetime measurement with $^4$He(K$^-$,
  $\pi^0$){\HH} reaction}
  \urlprefix\url{http://j-parc.jp/researcher/Hadron/en/pac_2001/pdf/P77_2020-01.pdf}

\bibitem{JLAB-C12-19-002-proposal}
Gogami T, Nakamura S, Garibaldi F, Markowitz P, Reinhold J, Tang L and Urciuoli
  G Proposal to jlab pac48, c12-19-002: High accuracy measurement of nuclear
  masses of $\lambda$ hyperhydrogens
  \urlprefix\url{https://indico.jlab.org/event/394/contributions/6380/attachments/5334/6632/e12-19-002_JLab-PAC48_gogami_20200811.pdf}

\bibitem{PhysRevLett.114.232501}
Esser A, Nagao S, Schulz F, Achenbach P, Ayerbe~Gayoso C, B\"ohm R, Borodina O,
  Bosnar D, Bozkurt V, Debenjak L, Distler M~O, Fri\v{s}\'{c} I, Fujii Y,
  Gogami T, Hashimoto O, Hirose S, Kanda H, Kaneta M, Kim E, Kohl Y, Kusaka J,
  Margaryan A, Merkel H, Mihovilovi\v{c} M, M\"uller U, Nakamura S~N,
  Pochodzalla J, Rappold C, Reinhold J, Saito T~R, Sanchez~Lorente A,
  S\'anchez~Majos S, Schlimme B~S, Schoth M, Sfienti C, \v{S}irca S, Tang L,
  Thiel M, Tsukada K, Weber A and Yoshida K (A1 Collaboration) 2015 {\em Phys.
  Rev. Lett.\/} {\bf 114}(23) 232501
  \urlprefix\url{http://link.aps.org/doi/10.1103/PhysRevLett.114.232501}

\bibitem{Schulz2016149}
Schulz F, Achenbach P, Aulenbacher S, Bericic J, Bleser S, B{\"o}hm R, Bosnar
  D, Correa L, Distler M, Esser A, Fonvieille H, Fri\v{s}cic I, Fujii Y, Fujita
  M, Gogami T, Kanda H, Kaneta M, Kegel S, Kohl Y, Kusaka W, Margaryan A,
  Merkel H, Mihovilovic M, M\"uller U, Nagao S, Nakamura S, Pochodzalla J,
  Lorente A~S, Schlimme B, Schoth M, Sfienti C, \v{S}irca S, Steinen M,
  Takahashi Y, Tang L, Thiel M, Tsukada K, Tyukin A and Weber A 2016 {\em
  Nuclear Physics A\/} {\bf 954} 149 -- 160 ISSN 0375-9474 recent Progress in
  Strangeness and Charm Hadronic and Nuclear Physics
  \urlprefix\url{http://www.sciencedirect.com/science/article/pii/S037594741630001X}

\bibitem{Herrmann2021PhDthesis}
Herrmann P 2021 {\em {Vorarbeiten f\"ur genaue und pr\"azise Messungen leichter
  Hyperkernmassen mit der A1-Spektrometeranlage am MAMI}\/} Ph.D. thesis
  {Johannes Gutenberg-Universit{\"a}t Mainz}

\bibitem{KAISER2008159}
Kaiser K~H, Aulenbacher K, Chubarov O, Dehn M, Euteneuer H, Hagenbuck F, Herr
  R, Jankowiak A, Jennewein P, Kreidel H~J, Ludwig-Mertin U, Negrazus M,
  Ratschow S, Schumann S, Seidl M, Stephan G and Thomas A 2008 {\em Nuclear
  Instruments and Methods in Physics Research Section A: Accelerators,
  Spectrometers, Detectors and Associated Equipment\/} {\bf 593} 159--170 ISSN
  0168-9002
  \urlprefix\url{https://www.sciencedirect.com/science/article/pii/S0168900208007341}

\bibitem{Achenbach:2018WpProc}
Achenbach P, Bleser S, Pochodzalla J and Steinen M 2018 {\em Proceedings of
  XVII International Conference on Hadron Spectroscopy and Structure, PoS
  (Hadron2017)\/} vol 310 p 207
  \urlprefix\url{https://pos.sissa.it/310/207/pdf}

\bibitem{Eckert:2022G6}
Eckert P, Achenbach P, Aragon\'{e}s~Fontbot\'{e} M, Akiyama T, Distler M~O,
  Esser A, Geratz J, Hoek M, Itabashi K, Kaneta M, Klag P, Merkel H, Mizuno M,
  M{\"u}ller J, M{\"u}ller U, Nagao S, Nakamura S~N, Nakamura Y, Okuyama K,
  Pochodzalla J, Sfienti C, Spreckels R, Steinen M, Thiel M, Uehara K and
  Toyama Y 2022 {\em Proceedings of Particles and Nuclei International
  Conference 2021 {\textemdash} PoS(PANIC2021)\/} vol 380 p 201

\bibitem{KLAG2018147}
Klag P, Achenbach P, Biroth M, Gogami T, Herrmann P, Kaneta M, Konishi Y, Lauth
  W, Nagao S, Nakamura S, Pochodzalla J, Roser J and Toyama Y 2018 {\em Nuclear
  Instruments and Methods in Physics Research Section A: Accelerators,
  Spectrometers, Detectors and Associated Equipment\/} {\bf 910} 147--156 ISSN
  0168-9002
  \urlprefix\url{https://www.sciencedirect.com/science/article/pii/S0168900218312166}

\bibitem{PhysRevLett.80.5473}
Dambach S, Backe H, Doerk T, Eftekhari N, Euteneuer H, G\"orgen F, Hagenbuck F,
  Kaiser K~H, Kettig O, Kube G, Lauth W, Sch\"ope H, Steinhof A, Tonn T and
  Walcher T 1998 {\em Phys. Rev. Lett.\/} {\bf 80}(25) 5473--5476
  \urlprefix\url{https://link.aps.org/doi/10.1103/PhysRevLett.80.5473}

\bibitem{JY1993}
{HORIBA Scientific} 1993 {\em {Operators Manual: HR640}\/} Jobin Yvon,
  Longjumeau, France
  \urlprefix\url{http://www.horiba.com/fileadmin/uploads/Scientific/Documents/OSD/HR640.pdf}

\bibitem{HAMAMATSU:2019}
{Hamamatsu Photonics} 2019 {\em {Datasheet: Orca Fusion}\/} Hamamatsu, 812
  Joko-cho, Higashi-ku, Hamamatsu City, 431-3196, Japan
  \urlprefix\url{https://unicam.hu/files/mikroszkopos-kamerak/hamamatsu-kamerak/cmos-kamerak/orca-fusion-c14440-20up/ORCA_Fusion_Technical_Note.pdf}

\bibitem{NIST_ASD}
Kramida A, {Yu~Ralchenko}, Reader J and {and NIST ASD Team} 2015 {NIST Atomic
  Spectra Database(ver. 5.3), [Online]. Available:
  {\ttfamily{http://physics.nist.gov/asd}} [2016, Nov 16]. National Institute
  of Standards and Technology, Gaithersburg, MD.}

\end{thebibliography}
\end{document}